\documentclass[aps,prb,showpacs,twocolumn,amssymb,floats,epsfig]{revtex4}
\usepackage{epsfig,psfrag,subfigure}
\usepackage{graphicx,psfrag,subfigure}
\usepackage{color}
\usepackage{amssymb,amsbsy,amsmath}

\newcommand\beq{\begin{equation}}
\newcommand\eeq{\end{equation}}
\newcommand\bea{\begin{eqnarray}}
\newcommand\eea{\end{eqnarray}}
\newcommand\al{\alpha}

\newcommand\ga{\gamma}
\newcommand\de{\delta}
\newcommand\ep{\epsilon}
\newcommand\De{\Delta}
\newcommand\si{\sigma}

\newcommand\la{\lambda}
\newcommand\om{\omega}
\newcommand\ta{\theta}

\newcommand\non{\nonumber}
\newcommand\noi{\noindent}

\newcommand\ig{\includegraphics}
\newcommand\bib{\bibitem}

\begin{document}

\title{Floquet generation of Majorana end modes and topological invariants} 

\author{Manisha Thakurathi$^1$, Aavishkar A. Patel$^2$, Diptiman Sen$^1$, 
and Amit Dutta$^3$}
\affiliation{\small{
$^1$Centre for High Energy Physics, Indian Institute of Science, Bangalore
560 012, India \\
$^2$Department of Physics, Harvard University, Cambridge, Massachusetts 02138,
USA \\
$^3$Department of Physics, Indian Institute of Technology, Kanpur 208 016, 
India}}

\date{\today}

\begin{abstract}
We show how Majorana end modes can be generated in a one-dimensional system
by varying some of the parameters in the Hamiltonian periodically in time. 
The specific model we consider is a chain containing spinless electrons with 
a nearest-neighbor hopping amplitude, a $p$-wave superconducting term and 
a chemical potential; this is equivalent 
to a spin-1/2 chain with anisotropic $XY$ couplings between nearest neighbors 
and a magnetic field applied in the $\hat z$ direction. We show that varying 
the chemical potential (or magnetic field) periodically in time can produce 
Majorana modes at the ends of a long chain. We discuss two kinds of periodic 
driving, periodic $\de$-function kicks and a simple harmonic variation with 
time. We discuss some distinctive features of the end modes such as 
the inverse participation ratio of their wave functions and their Floquet 
eigenvalues which are always equal to $\pm 1$ for time-reversal symmetric
systems. For the case of periodic $\de$-function kicks, we use the 
effective Hamiltonian of a system with periodic boundary conditions to define 
two topological invariants. The first invariant is a well-known winding 
number while the second invariant has not appeared in the literature before. 
The second invariant is more powerful in that it always correctly predicts 
the numbers of end modes with Floquet eigenvalues equal to $+1$ and $-1$,
while the first invariant does not. We find that the number of end modes 
can become very large as the driving frequency decreases. We show that
periodic $\de$-function kicks in the hopping and superconducting terms can
also produce end modes. Finally, we study the effect of electron-phonon 
interactions (which are relevant at finite temperatures) and a random noise 
in the chemical potential on the Majorana modes.

\end{abstract}

\pacs{71.10.Pm, 03.65.Vf, 75.10.Pq}
\maketitle

\section{Introduction}
Topological phases of quantum matter have been extensively studied for 
several years~\cite{hasan,qi,fidkowski1}. Typically, these are phases which
have only gapped states in the bulk (which therefore do not contribute
at low temperatures to properties like transport) but have gapless states 
at the boundaries. (For three-, two- and one-dimensional systems, the 
boundaries are given by surfaces, edges and end points respectively). Further,
the number of species of gapless boundary modes is given by a topological 
invariant whose nature depends on the spatial dimensionality of the system and
the symmetries that it possesses, like spin rotation symmetry, particle-hole 
symmetry and time-reversal symmetry. The significance of a 
topological invariant is that it does not change if the system is perturbed 
(say, by impurities), as long as the bulk states remain gapped and 
the symmetry of the system is not changed by the perturbation. Examples of 
systems with topological phases include two- and three-dimensional
topological insulators, quantum Hall systems, and wires with $p$-wave 
superconductivity. 

Recently, there has been considerable interest in systems in which the
Hamiltonian varies with time in a periodic way which gives rise to some 
topological features~\cite{kita1,lind1,jiang,gu,kita2,lind2,trif,russo,basti1,
liu,tong,cayssol,rudner,basti2,tomka,gomez,dora,katan,kundu,basti3,schmidt,
reynoso,wu}.
Some of these papers have discussed boundary modes and topological 
invariants~\cite{kita1,lind1,jiang,trif,liu,tong,cayssol,rudner,kundu}.
Recently a photonic topological insulator has been demonstrated 
experimentally; a two-dimensional lattice of helical waveguides has been 
shown to exhibit topologically protected edge states~\cite{recht}.
However, the existence of topological invariants and the relation between 
them and the number of Majorana modes at the boundary seems to be unclear, 
particularly if 
the driving frequency is small~\cite{tong}. Further, the Majorana boundary 
modes are of two types (corresponding to eigenvalues of the Floquet operator 
being $+1$ or $-1$, as discussed below); it would be interesting to know how
the numbers of these two types of modes can be obtained from a
topological invariant. The effect of time-reversal symmetry breaking on the 
boundary modes have also not been studied in detail. In this paper, we address
all these questions for a one-dimensional model where both Majorana end modes 
and topological invariants can be numerically studied without great difficulty.

The plan of this paper is as follows. In Sec. II we introduce the system
of interest and review some of its properties. Our system is a tight-binding
model of spinless electrons with $p$-wave superconducting pairing and a 
chemical potential. By the Jordan-Wigner transformation~\cite{lieb}, this can 
be shown to be equivalent to a spin-1/2 $XY$ chain 
placed in a magnetic field pointing in the $\hat z$ direction. We discuss
the energy spectrum and the three phases that this model has when the 
Hamiltonian is time-independent. In Sec. III, we review the topological 
invariants which one-dimensional models with and without time-reversal 
symmetry have when periodic boundary conditions are imposed. In Sec. IV, 
we discuss our numerical method of studying the Floquet evolution and
the modes which appear at the ends of a system when the Hamiltonian 
varies with time in a periodic way. In Sec. V, we study what happens 
when one of the terms in the Hamiltonian (the chemical potential in the 
electron language or the magnetic field in the spin language) is given a 
periodic $\de$-function kick~\cite{stock}. We study the ranges of parameters 
in which Majorana end modes appear at the ends of an open system and various 
properties of these modes such as their number and Floquet eigenvalues. We 
then use the Floquet operator for a system with periodic boundary conditions 
to define two topological invariants. The first invariant is a winding number
which gives the total number of end modes. The second invariant appears 
to be new; we find that it correctly predicts the numbers of end modes 
with Floquet eigenvalues equal to $+1$ and $-1$ separately. 
We find that end modes can either appear or disappear as the 
driving frequency is varied, and our second topological invariant predicts 
where this occurs. For a special choice of parameters, we are able to find 
analytical expressions for the wave functions of the Majorana end modes and to 
confirm that the second topological invariant correctly gives the
numbers of end modes with Floquet eigenvalues equal to $\pm 1$.
The effect of time-reversal symmetry breaking on the end modes is studied; 
we find that that the end modes may survive but they are no longer of the 
Majorana type. In Sec. VI, we briefly study what happens if the hopping 
amplitude and superconducting term are given periodic $\de$-function kicks. 
We show that the effect of this on the Majorana end modes is quite different 
from the case in which the chemical potential is given $\de$-function kicks.
In Sec. VII, we consider the case in which the chemical potential varies in 
time in a simple harmonic way, and we show that the wave function of the 
end modes can change significantly with time. In Sec. VIII, we study 
the effects of some aperiodic perturbations such as electron-phonon 
interactions and noise on the Majorana end modes. We summarize our main 
results and point out some directions for future work in Sec. IX.

\section{The Model}
We consider a lattice model of spinless electrons with
a nearest-neighbor hopping amplitude $\ga$, a $p$-wave superconducting 
pairing $\De$ between neighboring sites, and a chemical potential $\mu$. For a 
finite and open chain with $N$ sites, the Hamiltonian takes the form
\bea H &=& \sum_{n=1}^{N-1} [ \ga ( f_n^\dag f_{n+1} + f_{n+1}^\dag f_n ) 
+ \De ( f_n f_{n+1} + f_{n+1}^\dag f_n^\dag )] \non \\
&& - \sum_{n=1}^N \mu (2f_n^\dag f_n - 1), \label{ham1} \eea
where $\ga$, $\De$ and $\mu$ are all real; we may assume that $\ga > 0$ 
without loss of generality. In this section we will assume that all these
parameters are 
time-independent. The operators $f_n$ in Eq.~\eqref{ham1} satisfy the usual 
anticommutation relations $\{ f_m, f_n \} =0$ and $\{ f_m, f_n^\dag \} = 
\de_{mn}$. (We will set both Planck's constant $\hbar$ and the lattice spacing
equal to 1 in this paper). We introduce the Majorana operators 
\beq a_{2n-1} = f_n + f_n^\dag ~~~~{\rm and}~~~~ a_{2n} = i (f_n - f_n^\dag),
\label{majo} \eeq
for $n=1,2,\cdots,N$. We can check that these
are Hermitian operators satisfying $\{ a_m, a_n \} = 2 \de_{mn}$.
In terms of these operators, Eq.~\eqref{ham1} takes the form
\bea H &=& i ~ \sum_{n=1}^{N-1} ~[~ J_x a_{2n} a_{2n+1} ~-~ J_y a_{2n-1} 
a_{2n+2}] \non \\
& & +~ i ~ \sum_{n=1}^N ~\mu a_{2n-1} a_{2n}, \non \\
J_x &=& \frac{1}{2} (\ga - \De) ~~~~{\rm and}~~~~ J_y = \frac{1}{2} 
(\ga + \De). \label{ham2} \eea
Note that the Hamiltonian is invariant under the parity transformation 
$\cal P$ corresponding to a reflection of the system about its mid-point, 
i.e., $a_{2n} \to (-1)^n a_{2N+1-2n}$ and $a_{2n+1} \to a_{2N-2n}$.

We can map the above system to a spin-1/2 $XY$ chain placed in a magnetic 
field pointing in the $\hat z$ direction. We define the Jordan-Wigner 
transformation from $N$ spin-1/2's to $2N$ Majorana operators~\cite{lieb},
\bea a_{2n-1} &=& \left( \prod_{j=1}^{n-1} \si_j^z \right) ~\si_n^x, \non \\
a_{2n} &=& \left( \prod_{j=1}^{n-1} \si_j^z \right) ~\si_n^y, \label{jw} \eea
where the $\si_n^a$ denote the Pauli matrices at site $n$, and 
$n=1,2,\cdots,N$. Eq.~\eqref{ham2} can then be rewritten as
\beq H ~=~ - \sum_{n=1}^{N-1} ~[~ J_x \si_n^x \si_{n+1}^x ~+~ J_y \si_n^y
\si_{n+1}^y] ~- \sum_{n=1}^N ~\mu \si_n^z. \label{ham3} \eeq
In all our numerical calculations, we will set $\ga = - \De$; this implies 
that $J_y = 0$ and $J_x = \ga$, so that our system will be equivalent to an 
Ising model (with interaction $J_x$) in a transverse magnetic field $\mu$.

The system discussed above is time-reversal symmetric. The time-reversal
transformation involves complex conjugating all objects, including $i \to - 
i$. With the usual convention for the Pauli matrices, Eq.~\eqref{jw} implies 
that
\beq a_{2n} \to -~ a_{2n} ~~~~{\rm and}~~~~ a_{2n+1} \to a_{2n+1}. 
\label{trs} \eeq
Hence Eq.~\eqref{ham2} is time-reversal symmetric.

The energy spectrum of this system in the bulk can be found by considering 
a chain with periodic boundary conditions. We define the Fourier transform 
$f_k = \frac{1}{\sqrt N}~ \sum_{n=1}^N f_n e^{ikn}$, where the momentum $k$ 
goes from $-\pi$ to $\pi$ in steps of $2\pi/N$. Then Eq.~\eqref{ham1}) can 
be written in momentum space as
\bea H &=& 2 (\ga - \mu) f_0^\dag f_0 ~+~ 2 (- \ga - \mu) f_\pi^\dag f_\pi
\non \\
&& +~ \sum_{0 < k < \pi} ~\left( \begin{array}{cc}
f_k^\dag & f_{-k} \end{array} \right) ~h_k ~\left( \begin{array}{c}
f_k \\
f_{-k}^\dag \end{array} \right), \non \\
h_k &=& 2(\ga \cos k - \mu) ~\tau^z ~+~ 2 \De \sin k ~\tau^y, \label{hk} \eea
where the $\tau^a$ are Pauli matrices denoting pseudo-spin. 
The dispersion relation follows from Eq.~\eqref{hk} and is given 
by~\cite{gott1,gott2}
\beq E_k ~=~ \sqrt{4(\ga \cos k - \mu)^2 ~+~ 4 \De^2 \sin^2 k}.
\label{disp} \eeq

\begin{figure}[htb] \ig[width=3.4in]{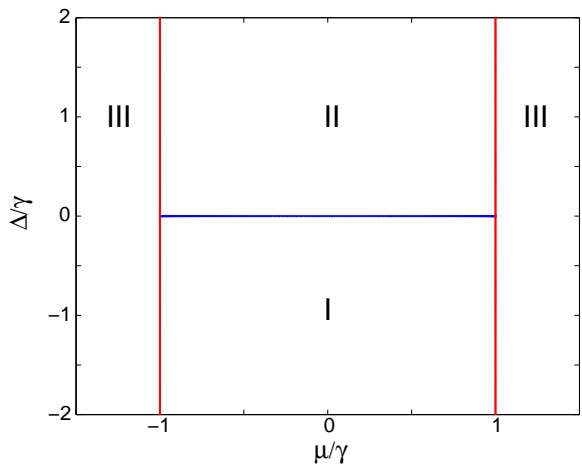}
\caption[]{(Color online) Phase diagram of the model in Eq.~\eqref{ham1} as 
a function of $\mu/\ga$ and $\De/\ga$. Phases I and II are topological 
while III is non-topological.} \label{flo0} \end{figure} 

Depending on the values of $\ga$, $\De$ and $\mu$, the system has three 
phases where $E_k$ is non-zero for all values of $k$~\cite{gott1,gott2}. 
The phase diagram is shown in Fig.~\ref{flo0}. Phase I lies in the region 
$\De /\ga < 0$ and $- 1 < \mu/\ga < 1$. In this phase, a long and open chain 
has a zero energy Majorana mode at the left (right) end in which $a_m$ is
non-zero only if $m$ is odd (even). This can be seen by considering the
extreme case $J_x > 0$ and $J_y = \mu = 0$ in Eq.~\eqref{ham2}. Then that 
Hamiltonian is independent of $a_1$ at the left end and $a_{2N}$
at the right end; hence we have zero energy modes corresponding to these
two operators. In the spin-1/2 language of Eq.~\eqref{ham3}, phase I 
corresponds to long-range ferromagnetic order of $\si^x$. Next, phase II lies 
in the region $\De /\ga > 0$ and $- 1 < \mu/\ga < 1$; here a long and open 
chain has a zero energy Majorana mode at the left (right) end in which $a_m$ 
is non-zero only if $m$ is even (odd). In the spin-1/2 language, this phase 
corresponds to long-range ferromagnetic order of $\si^y$. Finally, phase III 
consists of the two regions with $\mu/\ga < -1$ and $\mu/\ga > 1$. In this 
phase, there are no zero energy Majorana modes at either end of an open chain.
In the spin language, this is a paramagnetic phase with no long-range order.
The three phases are separated from each other by quantum critical lines
where the energy $E_k$ vanishes for some values of $k$. The critical lines
are given by $\mu/\ga = \pm 1$ for all values of $\De$, and $-1 \le \mu/\ga \le
1$ for $\De = 0$. We will see in the next section that the three phases 
can be distinguished from each other by a topological invariant which is 
given by a winding number.

\section{Topological Invariants for a Time-independent Hamiltonian}
In this section, we review the meaning of a topological phase 
and the topological invariants which exist for a one-dimensional system 
with a time-independent Hamiltonian which may or may not have time-reversal 
symmetry~\cite{gott2}. This discussion will be useful for Sec. V where we 
will study if similar topological invariants exist for a system in which 
the Hamiltonian varies periodically with time.

We begin by considering a general Hamiltonian which
is quadratic in terms of Majorana fermions,
\beq H ~=~ i \sum_{m,n=1}^{2N} ~a_m M_{mn} a_n, \label{ham4} \eeq
where $M$ is a real antisymmetric matrix; hence $iM$ is Hermitian.
We can show that the non-zero eigenvalues of $iM$ come in pairs
$\pm \la_j$ (where $\la_j > 0$), and the corresponding eigenvectors are 
complex conjugates of each other, $x_j$ and $x^*_j$. This follows
because $iM x_j = \la_j x_j$ implies $iM x_j^* = - \la_j x^*_j$.
The zero eigenvalues must be even in number and their eigenvectors can
be chosen to be real. This is because $i M x_j = 0$ implies $iM x^*_j = 0$,
and we can then choose the eigenvectors to be the real combinations
$x_j + x^*_j$ and $i(x_j - x^*_j)$.

Given the time-reversal transformation in Eq.~\eqref{trs}, we see that
the Hamiltonian in Eq.~\eqref{ham4} will have time-reversal symmetry if 
the matrix elements $M_{mn}$ are zero whenever both $m$ and $n$ are even
or both are odd. Further, let us assume that the system is translation
invariant and has periodic boundary conditions so that $M_{mn}$ is only
a function of $m-n$ modulo $2N$. Defining the Dirac fermions $f_n$ using
Eq.~\eqref{majo}, we then find that the Hamiltonian will have the form
given in Eq.~\eqref{hk}, with~\cite{gott2}
\beq h_k ~=~ a_{2,k} ~\tau^y ~+~ a_{3,k} ~\tau^z, \label{a23k} \eeq
where $a_{2/3,k}$ are some real and periodic functions of $k$. 
The corresponding dispersion is then given by $E_k = \sqrt{a_{2,k}^2 + 
a_{3,k}^2}$. Although Eq.~\eqref{hk} defines $a_{2/3,k}$ only for 
$0 \le k \le \pi$, it is convenient to analytically continue 
these definitions to the entire range $-\pi \le k \le \pi$. Next we map 
$h_k$ to the vector $\vec V_k = a_{2,k} \hat y + a_{3,k} \hat z$ in the $y-z$ 
plane. Let us define the angle $\phi_k = \tan^{-1} (a_{3,k}/a_{2,k})$ made by 
the vector $\vec V_k $ with respect to the $\hat z$ axis. Following 
Refs.~[\onlinecite{niu,tong}], we now define a winding number by 
following the change in $\phi_k$ as we go around the Brillouin zone, i.e.,
\beq W ~=~\int_{-\pi}^{\pi} ~\frac{dk}{2\pi} ~\frac{d\phi_k}{dk}. 
\label{wind} \eeq
This can take any integer value and is a topological invariant, namely, it
does not change under small changes in $h_k$ unless $h_k$ happens to pass
through zero for some value of $k$ in which case the winding number becomes
ill-defined; this can only happen if the energy $E_k = 0$ at some value of 
$k$ which means that the bulk gap is zero. In a gapped phase, therefore,
Eq.~\eqref{wind} defines a $Z$-valued topological invariant. We call a phase
topological if $W \ne 0$; such a phase will have $W$ zero energy Majorana 
modes at each end of long chain~\cite{gott2}. If $W = 0$, the phase is 
non-topological and does not have any Majorana end modes.

We can now look at the three phases discussed after Eq.~\eqref{disp}.
We discover, by taking appropriate limits (like $\mu \ll \ga, ~\De$ or
$\mu \gg \ga, ~\De$) that the winding number takes the values $-1$, $+1$ and
$0$ in phases I, II and III respectively.

Next, we note that if time-reversal symmetry breaking terms were present in 
the Hamiltonian in \eqref{ham4}, terms proportional to $\tau^x$ and the 
identity matrix $I$ will appear in $h_k$ in addition to terms proportional to 
$\tau^y$ and $\tau^z$. Then as $k$ goes from $-\pi$ to $\pi$, $h_k$ will 
generate a closed curve in three or four dimensions 
instead of only two dimensions, and it would not be possible to define a 
winding number as a topological invariant. However, it turns out that one can 
define a $Z_2$-valued topological invariant in that case~\cite{kitaev,gott2}.
We find that at $k=0$ and $\pi$, $h_k$ only has a component along $\tau^z$;
this is essentially because $k=-k$ in those two cases, hence terms
proportional to $\tau^x$, $\tau^y$ and $I$ cannot appear in $h_k$. Let us 
denote $h_0 = g_0 \tau^z$ and $h_\pi = g_\pi \tau^z$. Assuming that we are in 
a gapped phase, so that $h_k \ne 0$ for all values of $k$, the $Z_2$-valued 
topological invariant is defined as $\nu = sgn (g_0 g_\pi)$ (here $sgn$ 
denotes the signum function). If $\nu = -1$, the phase is topological and has 
one zero energy Majorana mode at each end of a long chain, but if $\nu = 1$, 
the phase is non-topological and does not have any Majorana end modes.

Finally, we can ask what would happen if one considered a time-reversal 
symmetric system which is in a topological phase with winding number $W$
(and hence has $W$ zero energy Majorana modes at each end of a long chain),
and introduced a weak time-reversal breaking term in the Hamiltonian.
Generally, what happens is that pairs of end modes move away from zero 
energy to energies $\pm E$; the number of modes which remain at zero energy 
(and hence are Majorana modes) is $1$ if $W$ is odd and zero if $W$ is even.
Thus, the $Z$-valued invariant $W$ would reduce to the $Z_2$-valued 
invariant $\nu$ as $\nu = (-1)^W$.

\section{Floquet Evolution}
We will now study what happens when the Hamiltonian varies
periodically in time, namely, the matrix $M$ in Eq.~\eqref{ham4} changes with 
time as $M(t)$ such that $M(t+T)=M(t)$, where $T$ denotes the time period.

We consider the Heisenberg operators $a_n (t)$. These satisfy the equations
\beq \frac{da_n(t)}{dt} ~=~ i ~[H(t),a_n (t)]. \eeq
Given that $H(t) = i \sum_{mn} a_m (t) M_{mn} (t) a_n (t)$, we obtain
\beq \frac{da_m(t)}{dt} ~=~ 4 \sum_{n=1}^{2N} ~ M_{mn} (t) ~a_n (t). \eeq
If $a$ denotes the column vector $(a_1a_2,\cdots,a_{2N})^T$ and $M$ 
denotes the matrix $M_{mn}$, we can write the above equation as 
$da(t)/dt = 4 M(t) a(t)$. The solution of this is given by
\bea a (t) &=& U(t,0) ~a(0), \non \\
{\rm where}~~ U(t_2,t_1) &=& {\cal T} e^{4 \int_{t_1}^{t_2} dt M(t)}, \eea
and $\cal T$ denotes the time-ordering symbol. The time evolution operator 
$U(t_1,t_2)$ is a unitary (in fact, real and orthogonal) matrix which can be 
numerically computed given the form of $M(t)$. It satisfies the properties 
$U(t_2,t_1) = U^{-1} (t_1,t_2)$ and $U(t_3,t_1) = U(t_3,t_2) U(t_2,t_1)$. 

If $M(t)$ varies with a time period $T$, we will call 
$U(T,0)$ the Floquet operator. The eigenvalues of $U(T,0)$ are given by 
phases, $e^{i\ta_j}$, and they come in complex conjugate pairs if
$e^{i\ta_j} \ne 1$. This is because $U(T,0) \psi_j = e^{i\ta_j} \psi_j$ 
implies that $U(T,0) \psi_j^* = e^{-i\ta_j} \psi_j^*$. 
For eigenvalues $e^{i\ta_j} = \pm 1$ (these eigenvalues may, in principle, 
appear with no degeneracy), the eigenvectors can be chosen to be real; one 
can show this using an argument similar to the one given above for zero 
eigenvalues of the matrix $iM$.

In Secs. V and VII, we will consider two kinds of periodic driving of the 
chemical potential $\mu (t)$ with a time period $T$, namely, periodic 
$\de$-function kicks~\cite{stock} and a simple harmonic variation with time.
In Sec. VI, we will consider what happens if the hopping amplitude and
superconducting term are given periodic $\de$-function kicks.
In each case, we will look for eigenvectors of $U(T,0)$ which are localized 
near the ends of the chain. Before discussing the specific results in the
next three sections, let us describe our method of finding Majorana end 
modes and some of their general properties. 

A convenient numerical method for finding eigenvectors of $U(T,0)$ which 
are localized at the ends is to look at the inverse participation ratio 
(IPR). We assume that the eigenvectors, denoted as $\psi_j$, are 
normalized so that $\sum_{m=1}^{2N} |\psi_j (m)|^2 = 1$ for each value of
$j$; here $m=1,2,\cdots,2N$ labels the components of the eigenvector. 
We then define the IPR of an eigenvector as $I_j = \sum_{m=1}^{2N} |\psi_j 
(m)|^4$. If $\psi_j$ is extended equally over all sites so that 
$|\psi_j (m)|^2 = 1/(2N)$ for each $m$, then $I_j = 1/(2N)$;
this will approach zero as $N \to \infty$. But if $\psi_j$ is localized over
a distance $\xi$ (which is of the order of the decay length of the eigenvector
and remains constant as $N \to \infty$), then we will have $|\psi_j (m)|^2 
\sim 1/\xi$ in a region of length $\xi$ and $\sim 0$ elsewhere; then we have 
$I_j \sim 1/\xi$ which will remain finite as $N \to \infty$. If $N$ is 
sufficiently large, a plot of $I_j$ versus $j$ will be able to distinguish 
between states which are localized (over a length scale $\ll N$) and states 
which are extended. Once we find a state $j$ for which $I_j$ is significantly 
larger than $1/(2N)$ (which is the value of the IPR for a completely extended 
state), we look at a plot of the probabilities $|\psi_j (m)|^2$ versus $m$
to see whether it is indeed an end state. Finally, we check if the form
of $|\psi_j (m)|^2$ and the value of IPR remain unchanged if $N$ is increased. 

In all the periodic driving protocols discussed in Secs. V, VI and VII, we 
find, for certain ranges of the parameter values, that $U(T,0)$ has one or more
pairs of eigenvectors with substantial values of the IPR. For each such pair, 
we find that the corresponding Floquet eigenvalues are complex conjugates 
of each other and they are both close to 1 (or $-1$); the two eigenvalues 
approach 1 (or $-1$) as we increase the system size $N$ keeping all the other 
parameters the same. Let us denote the corresponding eigenvectors by $\psi_1
(m)$ and $\psi_2(m)$, where $m=1,2,\cdots,2N$. In the limit that $N \to 
\infty$ and the eigenvalues approach 1 (or $-1$), any linear combination of 
$\psi_1$ and $\psi_2$ will also be an eigenvector of $U(T,0)$ with the same 
eigenvalue. In that limit, suppose that we find that the probabilities of the 
two orthogonal linear combinations, given by $|(\psi_1(m) \pm \psi_2(m))|^2$, 
are peaked close to $m=1$ and $2N$, and that they decay as $m$ moves away 
from 1 or $2N$. We can then interpret these linear combinations as edge 
states produced by the time-dependent chemical potential. The deviation of 
the two Floquet eigenvalues from 1 (or $-1$) is a measure of the 
tunneling between the two edge states. The larger the tunneling, the greater 
is the deviation of the eigenvalues from $\pm 1$; this, in turn, implies that 
the two edge states decay less rapidly as we go away from the ends of the 
chain since a slower decay increases the tunneling between the two states.

The situation discussed in the previous paragraph is similar in some respects 
to the problem of a time-independent double well potential in one dimension 
which is reflection symmetric about one point, say, $x=0$. Then the eigenstates
of the Hamiltonian are simultaneously eigenstates of the parity operator.
The lowest energy states in the parity even and parity odd sectors differ
in energy by an amount which depends on the tunneling amplitude between the 
two wells; the corresponding wave functions, denoted by $\psi_+$ and $\psi_-$, 
are symmetric and antisymmetric combinations of wave functions which are 
localized in the two wells separately. In the limit that the tunneling 
amplitude goes to zero, the two states become degenerate in energy; further, 
the linear combinations $\psi_+ \pm \psi_-$ describe states which are 
localized in the two separate wells. In our Floquet problem, the two end 
states with opposite parity have complex conjugate eigenvalues of $U(T,0)$ 
given by $e^{\pm i \ta}$. In the limit that $N \to \infty$ and the tunneling 
between the two states goes to zero, the eigenvalues of $U(T,0)$ must become 
degenerate; this can only happen if $e^{\pm i \ta}$ approach either $+1$ or 
$-1$.

Finally, after finding the end modes, we check if their wave functions are 
real in the limit of large $N$. We call the end modes Majorana if they satisfy
three properties: their Floquet eigenvalues must be equal to $\pm 1$, they 
must be separated by a finite gap from all the other eigenvalues, and their 
wave functions must be real.

\section{Periodic $\de$-function Kicks in Chemical Potential}
In this section, we consider the case where the chemical potential 
is given $\de$-function kicks periodically in time. One reason for choosing 
to consider periodic kicks is that this is known to produce interesting
effects in quantum systems such as dynamical localization~\cite{stock}. 
We will also see that this system is considerably easier to study both 
numerically and analytically than the case of a simple harmonic 
time-dependence which will be discussed in Sec. VII.

We begin by taking the chemical potential in Eq.~\eqref{ham2} to be 
of the form
\beq \mu (t) ~=~ c_0 ~+~ c_1 \sum_{n=-\infty}^\infty \de (t - nT), 
\label{ht1} \eeq
where $T=2\pi/\om$ is the time period and $\om$ is the driving
frequency. Using Eq.~\eqref{trs}, we note that 
this system has time-reversal symmetry: $H^* (-t) = H(t)$ for all values
of $t$. (In general, we say that a system has time-reversal symmetry if we 
can find a time $t_0$ such that $H^* (t_0 - t) = H(t)$ for all $t$, and does 
not have time-reversal symmetry if no such $t_0$ exists). As discussed below, 
we numerically compute the operator $U(T,0)$ for various 
values of the parameters $\ga$, $\De$, $c_0$, $c_1$, $\om$ and the system
size $N$. We then find all the eigenvalues and eigenvectors of $U(T,0)$.
Since the system is invariant under parity, $\cal P$, one can choose the
eigenvectors of $U(T,0)$ to also be eigenvectors of $\cal P$.

The Floquet operator for a periodic 
$\de$-function kick can be written as a product of 
two terms: an evolution with a constant chemical potential $c_0$ for time $T$ 
followed by an evolution with a chemical potential $c_1 \de (t-T)$. Namely,
\beq U(T,0) ~=~ e^{4 M_1} ~e^{4 M_0 T}, \label{flo1x} \eeq 
where $M_{0/1}$ are $(2N)$-dimensional antisymmetric matrices whose non-zero 
matrix elements can be found using Eqs.~\eqref{ham2} and \eqref{ham4}:
\bea (M_0)_{2n+1,2n} &=& - ~(M_0)_{2n,2n+1} ~=~ - \frac{1}{4} (\ga - \De), 
\non \\
(M_0)_{2n-1,2n+2} &=& - ~(M_0)_{2n+2,2n-1} ~=~ - \frac{1}{4} (\ga + \De), 
\non \\
(M_0)_{2n-1,2n} &=& - ~(M_0)_{2n,2n-1} ~=~ \frac{c_0}{2}, \non \\
(M_1)_{2n-1,2n} &=& - ~(M_1)_{2n,2n-1} ~=~ \frac{c_1}{2}, \label{mt1} \eea
for an appropriate range of values of $n$. However, in order to make the 
time-reversal symmetry more transparent, it turns out to be more convenient 
to use the symmetrized expression 
\beq U(T,0) ~=~ e^{2 M_1} ~e^{4 M_0 T} ~e^{2 M_1}. \label{flo2x} \eeq
It is easy to show that the Floquet operators in Eqs.~\eqref{flo1x} and 
\eqref{flo2x} have the same eigenvalues, while their eigenvectors are related
by a unitary transformation. We will see below that the symmetrized form in 
Eq.~\eqref{flo2x} leads to some simplifications when we derive an effective 
Hamiltonian and a topological invariant.

\begin{figure}[htb] \ig[width=3.4in]{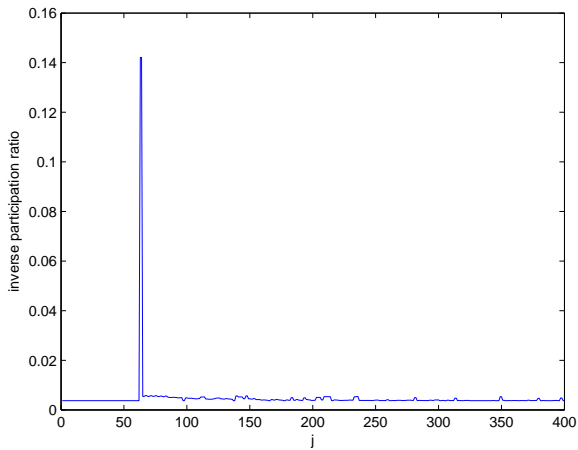}
\caption[]{(Color online) IPRs of different eigenvectors of the Floquet 
operator for a 200-site system with a periodic $\de$-function kick with 
$\ga =1$, $\De = - 1$, $c_0=2.5$, $c_1=0.2$ and $\om=12$. The two 
eigenvectors with the largest IPRs both have an IPR equal to $0.142$ and 
Floquet eigenvalue equal to $-1$.} \label{flo1} \end{figure}

\begin{figure}[htb] \ig[width=3.4in]{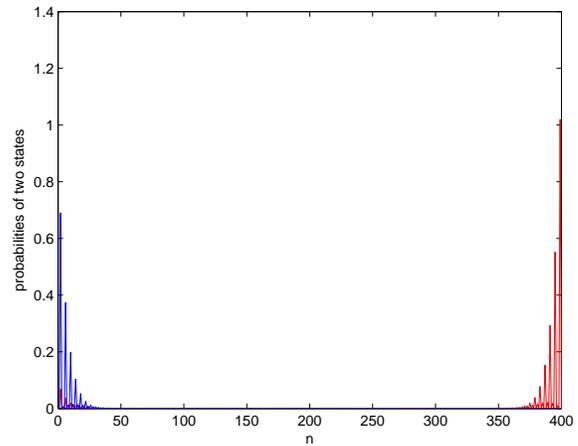}
\caption[]{(Color online) Majorana end states for a 200-site system with a 
periodic $\de$-function kick with $\ga = 1$, $\De = - 1$, $c_0=2.5$, $c_1=0.2$
and $\om=12$. These states correspond to the two eigenvectors with the largest 
IPRs in Fig.~\ref{flo1}.} \label{flo2} \end{figure}

We now consider a 200-site system (hence with a $400$-dimensional
Hamiltonian) with $\ga =1$, $\De = -1$, $c_0=2.5$, $c_1=0.2$ and $\om = 12$. 
Fig.~\ref{flo1} shows the IPRs of the different eigenvectors. Two of the 
IPRs clearly stand out with a value of $0.142$ each. We find that they both 
have Floquet eigenvalue $e^{i\ta} = -1$, and the value of $\ta=\pi$ is 
separated by a gap of $0.148$ from the values of $\ta$ for 
all the other eigenvalues. The 
corresponding eigenvectors are localized at the two ends of the system and 
are real; the corresponding probabilities are shown in Fig.~\ref{flo2}. The 
state at the left end has non-zero $a_m$ only if $m$ is even, while the state 
at the right end has non-zero $a_m$ only for $m$ odd. It is important to
note that the periodic driving has produced Majorana end modes even though
for the parameter values given above, $\mu (t) \ge \ga$ at all values of $t$ 
according to Eq.~\eqref{ht1}, i.e., even though the corresponding 
time-independent system lies at all times in phase III (discussed in the 
paragraph after Eq.~\eqref{disp}) which is a non-topological phase.

\begin{figure}[htb] \ig[width=3.4in]{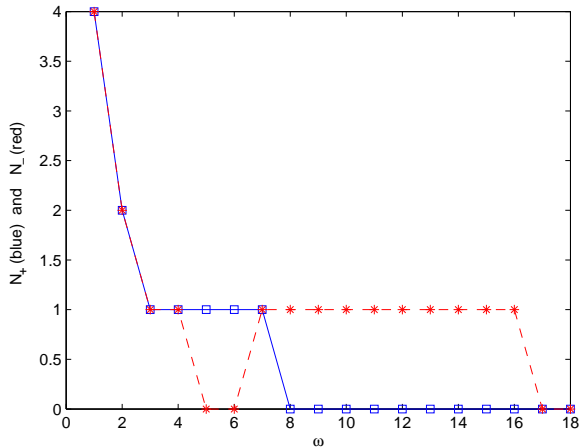}
\caption[]{(Color online) Plot of the number of end states versus $\om$ for a 
200-site system, with Floquet eigenvalues $+1$ ($N_+$, blue squares) and $-1$ 
($N_-$, red stars), for a periodic $\de$-function kick with $\ga =1$, $\De = 
-1$, $c_0=2.5$ and $c_1=0.2$.} \label{flo3} \end{figure}

We now vary $\om$ to see how many Majorana end modes there are at each end 
of the system, and, more specifically, how many of these modes have Floquet
eigenvalues equal to $\pm 1$. We denote the number of 
eigenvalues lying near $+1$ and $-1$ by the integers $N_+$ and $N_-$ 
respectively. Fig.~\ref{flo3} shows a plot of $N_\pm$ versus $\om$ for a 
200-site system with $\ga =1$, $\De = -1$, $c_0=2.5$ and $c_1=0.2$. 
(We have checked that these eigenvalues are separated from all the 
other eigenvalues by a gap which remains finite as $N$ becomes large).
We see that although the number of end modes is not a monotonic function of 
$\om$, the number generally increases as $\om$ decreases. The reason for 
this will become clear below.

\subsection{Topological Invariants}
We saw above that there are a number of Majorana end modes, which can be
further separated into $N_\pm$ depending on whether the Floquet eigenvalues 
$e^{i\ta}$ lie near $+1$ or $-1$. Further, the eigenvalues $\ta = 0$ and 
$\pi$ are separated from all the other eigenvalues by a gap which remains 
finite as $N \to \infty$. We then expect the integers $N_+$ 
and $N_-$ to be topological invariants, i.e., they will not change under 
small changes in the various parameters of the system. The only way in which 
these integers can change is if the eigenvalue gap closes and reopens as 
we vary the system parameters.

We therefore look for a topological invariant for this time-dependent
problem~\cite{kita1,jiang,trif,tong,rudner}. Interestingly, we will 
discover that we can define a topological invariant in two different ways:
one is a winding number which only gives the total number of Majorana modes
at each end of a chain, while the other also gives the individual values of 
$N_+$ and $N_-$ which are the numbers of end modes with Floquet eigenvalues 
equal to $+1$ and $-1$. 

To define the topological invariants, we consider
a system with periodic boundary conditions. Then the system is translation 
invariant and the momentum $k$ is a good quantum number; the system decomposes
into a sum of subsystems labeled by different values of $k$ lying in the 
range $[0,\pi]$. For each value of $k$, we define a Floquet operator 
$U_k (T,0)$ which is a $2 \times 2$ unitary matrix. Using Eqs.~\eqref{hk}, 
\eqref{ht1} and \eqref{flo2x}, we find that 
\beq U_k (T,0) = e^{ic_1 \tau^z} e^{-i2T[(\ga \cos k -c_0) \tau^z + \De \sin k
\tau^y]} e^{ic_1 \tau^z}, \label{uk1} \eeq
where we take $k$ to lie in the full range $- \pi \le k \le \pi$.

Let us assume that $2c_1/\pi$ is not equal to an integer and $\De \ne 0$.
We now prove an interesting fact about $U_k (T,0)$, namely, that it can 
be equal to $\pm I$ only if $k=0$ or $\pi$ and if $T$ is given by a discrete 
set of values. First, given the above conditions on $c_1$ and $\De$, we can 
show that $U_k (T,0) \ne \pm I$ for any value of $k \ne 0$ or $\pi$. Next, if 
$k = 0$ or $\pi$, we see from Eq.~\eqref{uk1} that $U_k (T,0) \ne \pm I$ unless
$2T (c_0 \pm \ga) + 2c_1 = n \pi$, i.e., unless $\om = 2\pi/T$ satisfies
\beq \om ~=~ \frac{4\pi (c_0 \pm \ga)}{n \pi - 2c_1} \label{omeq} \eeq
for some integer value of $n$. The $\pm$ sign in Eq.~\eqref{omeq} corresponds
to $k=\pi$ and $0$ respectively. Eq.~\eqref{omeq} holds only for a discrete 
set of values $\om$. For all other values of $\om$, therefore, $U_k (T,0)$ 
will not be equal to $\pm I$ for any value of $k$. This also means that
for all $k$, the Floquet eigenvalues (which are given by the eigenvalues of 
$U_k (T,0)$) will be separated by a gap from $\pm 1$. We are now ready to
define our topological invariants.

{\bf First topological invariant}:
Given Eq.~\eqref{uk1}, let us define an effective Hamiltonian $h_{eff,k}$ as 
\beq U_k (T,0) ~=~ e^{-ih_{eff,k}}. \label{heff1} \eeq
The structure of Eq.~\eqref{uk1} is such that $h_{eff,k}$ takes the form
\beq h_{eff,k} ~=~ a_{2,k} ~\tau^y ~+~ a_{3,k} ~\tau^z \label{heff2} \eeq
as in Eq.~\eqref{a23k}. (Indeed, this is the reason we choose the Floquet 
operator of the form given in Eq.~\eqref{flo2x} rather than in 
Eq.~\eqref{flo1x}). Note that Eqs.~(\ref{uk1}) and (\ref{heff1}) do not 
determine $h_{eff,k}$ uniquely. To define $h_{eff,k}$ uniquely, we impose 
the condition that the coefficients in Eq.~\eqref{heff2} satisfy
$0 < \sqrt{a_{2,k}^2 + a_{3,k}^2} < \pi$. (It is possible to impose this
if $U_k (T,0) \ne \pm I$; this will be true if $\om$ does not satisfy
Eq.~\eqref{omeq}). Given the form in Eq.~\eqref{heff2},
we can then compute a winding number $W$ as described in Eq.~\eqref{wind}.
 
We note in passing that the condition $0 < a_{2,k}^2 + a_{3,k}^2 < \pi$ 
implies that $h_{eff,k}$ can be mapped to a point on the surface of a 
sphere whose polar angles $(\al,\beta)$ are given by $\al = \sqrt{a_{2,k}^2 
+ a_{3,k}^2}$ and $\beta = \tan^{-1} (a_{3,k}/a_{2,k})$. As $k$ goes from 0 to 
$2\pi$, we obtain a closed curve which does not pass through the north and 
south poles. The integer $W$ can then be related to the winding number of 
this curve around either the north pole or the south pole. Note that the
winding numbers around the north and south pole are given by the same integer.

\begin{figure}[htb] \ig[width=3.4in]{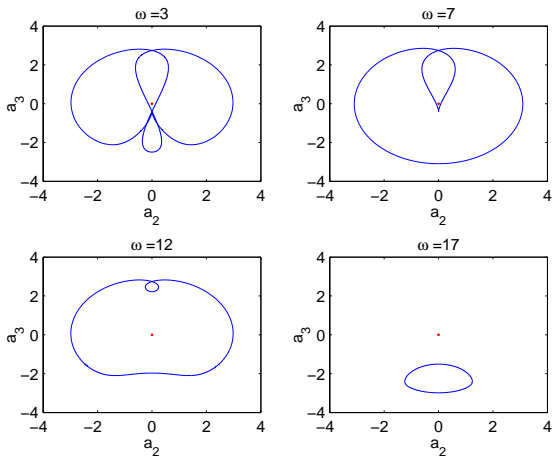}
\caption[]{(Color online) Closed curves in the $(a_{2,k}, a_{3,k})$ plane for 
$\om = 3, ~7, ~12$ and 17, for a 200-site system with $\ga =1$, $\De = -1$, 
$c_0=2.5$ and a periodic $\de$-function kick with $c_1 =0.2$. The 
corresponding winding numbers around the origin (the point $(0,0)$ shown by a 
red dot) are given by 2, 2, 1 and 0 respectively.} \label{flo4} \end{figure}

\begin{figure}[htb] \ig[width=3.4in]{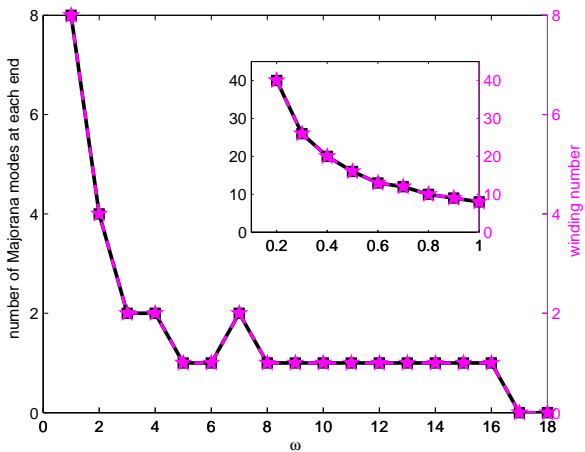}
\caption[]{(Color online) Comparison of the number of Majorana modes at each 
end of a 200-site system (black solid, $y$-axis on left) and the winding number
(magenta dashed, $y$-axis on right) as a function of $\om$ from 1 to 18, for 
$\ga =1$, $\De = -1$, $c_0=2.5$ and a periodic $\de$-function kick with $c_1
=0.2$. The inset shows a range of $\om$ from $0.2$ to 1 where there is a large
number of Majorana modes.} \label{flo5} \end{figure}

In Fig.~\ref{flo4}, we show the closed curves in the $(a_{2,k}, a_{3,k})$ 
plane for four values of $\om$ for a 200-site system 
with $\ga =1$, $\De = -1$, $c_0=2.5$ and a periodic $\de$-function kick 
with $c_1 =0.2$. For $\om = 3, ~7, ~12$ and 17, the winding numbers around 
the origin are given by 2, 2, 1 and 0 respectively.
These agree exactly with the number of Majorana modes at each end of an
open chain for those values of $\om$ as shown in Fig.~\ref{flo5}.

In Fig.~\ref{flo5}, we compare the number of Majorana modes at each end of a 
chain and the winding number as a function of $\om$, for a 200-site system 
with $\ga =1$, $\De = -1$, $c_0=2.5$ and $c_1=0.2$. In preparing that figure, 
we have considered only those values of $\om$ for which Eq.~\eqref{omeq} is 
not satisfied. We see that the number of end modes and the winding number 
completely agree in the range $0.2 \le \om \le 18$. 
Note that in the limit $\om \to \infty$,
i.e., $T \to 0$, Eq.~\eqref{uk1} becomes independent of $k$, and we therefore
obtain a single point in the $(a_{2,k}, a_{3,k})$ plane. This corresponds to 
a curve with zero winding number which is consistent with the observation
that there is a maximum value of $\om$ beyond which there are no 
Majorana end modes.

In Fig.~\ref{flo5}, we have not shown the number of Majorana end modes for 
$\om < 0.2$. For small $\om$, we see that the number of end modes increases. 
(We will make this more precise below).
However, it becomes more and more difficult to identify the end modes 
as $\om$ becomes small; we find that there are a large number of what appear 
to be end modes, but many of them have decay lengths which are not much smaller
than the system sizes that we have considered and their Floquet eigenvalues
differ slightly from $\pm 1$. Thus we have to go to very large system sizes 
to confirm if all of these are really Majorana end modes, i.e., if their 
Floquet eigenvalues approach $\pm 1$ and if these are separated from all 
other eigenvalues by a finite gap in the limit of infinite system size.

{\bf Second topological invariant}:
We observe that the momenta $k=0$ and $\pi$ play a special role since 
$U_k (T,0)$ can be equal to $\pm I$ at only those two values. Eq.~\eqref{uk1} 
shows that $U_0 (T,0) = e^{i\pi b_0 \tau^z}$ and $U_\pi (T,0) = e^{i\pi b_\pi 
\tau^z}$, where we choose $b_{0/\pi}$ in the simplest possible way, namely,
\bea b_0 &=& \frac{4(c_0 - \ga)}{\om} ~+~ \frac{2c_1}{\pi}, \non \\
b_\pi &=& \frac{4(c_0 + \ga)}{\om} ~+~ \frac{2c_1}{\pi}, \label{b0p} \eea
where $\om = 2\pi/T$. We now define a finite line segment, called $L_\om$, 
which goes from $b_0$ to $b_\pi$ in one dimension which we will call the 
$z$-axis. 

For $\om \to \infty$, the line $L_\om$ collapses to a single point given by 
$z=2c_1/\pi$. We have assumed earlier that this is not an integer. As $\om$ 
is decreased, $L_\om$ will move and also increase in size. For our system 
parameters $\ga = 1$, $c_0 = 2.5$ and $c_1 = 0.2$, we find that the {\it right}
end of $L_\om$, given by $b_\pi$ in Eq.~\eqref{b0p}, crosses the point $z=n$ 
with $n=1$ at some value of $\om$. At this point, we see from Eq.~\eqref{uk1} 
that the Floquet eigenvalue at $k=\pi$ is equal to $e^{in\pi} = -1$. We 
therefore expect that when $\om$ decreases a little more and $L_\om$ includes 
the point $z=1$, a Majorana mode will appear at each end of an open chain 
with the Floquet eigenvalue equal to $-1$. For our parameters, we therefore 
predict, by setting $b_\pi = 1$, that the first Majorana end 
mode will appear at $\om \simeq 
16.04$. This agrees well with Fig.~\ref{flo5} which shows that a Majorana end 
mode first appears in the range $16 \le \om \le 17$ and it has a Floquet 
eigenvalue equal to $-1$. As $\om$ is decreased further, the right end of 
$L_\om$ given by $b_\pi$ crosses the point $z=n$ with $n=2$ at another value 
of $\om$; Eq.~\eqref{uk1} then shows that the Floquet eigenvalue at 
$k=\pi$ is equal to $e^{in\pi} = 1$. As $\om$ is decreased a little more, 
$L_\om$ will include the point $z=2$, and we then expect that 
Majorana end modes will appear with the Floquet 
eigenvalue equal to $1$. For our parameters, $b_\pi =2$ occurs
at $\om \simeq 7.48$. This also agrees well with Fig.~\ref{flo5} which shows 
that a Majorana end mode appears in the range $7 \le \om \le 8$ with a 
Floquet eigenvalue equal to $1$. As $\om$ is decreased further, the {\it left}
end of $L_\om$, given by $b_0$ in Eq.~\eqref{b0p},
crosses the point $z=n$ with $n=1$ at some value of $\om$;
Eq.~\eqref{uk1} then shows that the Floquet eigenvalue at $k=0$ is equal to 
$e^{in\pi} = -1$. As $\om$ is decreased a little more, $L_\om$ no
longer includes the point $z=1$ and we expect that the Majorana end modes
with Floquet eigenvalue equal to $-1$ will {\it disappear}. For our
parameters, $b_0 = 1$ occurs at $\om = 6.88$. We see in Fig.~\ref{flo5}
that a Majorana end mode with Floquet eigenvalue equal to $-1$ disappears
in the range $6 \le \om \le 7$.

The general pattern is now clear. If $c_0 \pm \ga$ are both positive, the 
left and right ends of the line segment $L_\om$ will both move in the $+z$ 
direction as $\om$ decreases, i.e., as $T$ increases. Then a Majorana end 
mode with Floquet eigenvalue $(-1)^n$ will appear whenever the right end of 
$L_\om$ crosses a point $z=n$, while an end mode with Floquet eigenvalue 
$(-1)^n$ will disappear whenever the left end of $L_\om$ crosses $z=n$. These 
will happen, respectively, when $b_\pi$ and $b_0$ in Eq.~\eqref{b0p} become 
equal to an integer $n$. 

\begin{figure}[htb] \ig[width=3.4in]{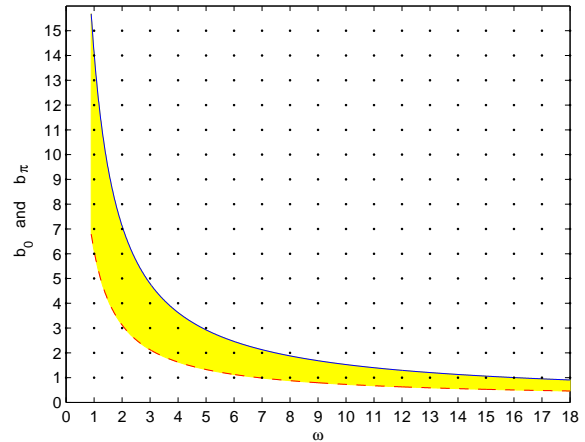}
\caption[]{(Color online) Plot of $b_0$ and $b_\pi$ as a function of $\om$ for 
a system with $\ga =1$, $\De = -1$, $c_0=2.5$ and a periodic $\de$-function 
kick with $c_1=0.2$. For each value of $\om$, the number of even and odd 
integers lying in the shaded region between $b_0$ and $b_\pi$ gives the number
of Majorana modes at each end of a chain with Floquet eigenvalues equal to 
$+1$ and $-1$ respectively.} \label{flo6} \end{figure}

The above arguments can be rephrased as follows. For any value of $\om$,
the number of points $z=n$ (where $n$ is an integer) which lie inside the line 
segment $L_\om$ is equal to the number of Majorana modes at each end of a 
chain. Further, the numbers of points with $n$ odd and even will give the 
numbers of end modes with Floquet eigenvalue equal to $-1$ and $1$ 
respectively. We have numerically verified these statements for all the 
values of $\om$ shown in Fig.~\ref{flo3}. In Fig.~\ref{flo6}, we show $b_0$ 
and $b_\pi$ (i.e., the left and right ends of $L_\om$) as functions of $\om$ 
for the parameters $\ga=1$, $\De = -1$, $c_0 =2.5$ and $c_1 = 0.2$. The
Majorana end modes correspond to the integers lying within the shaded region.

It is clear that the numbers of odd and even integers lying inside $L_\om$ 
are topological invariants since these numbers do not change for small 
changes of the system parameters. These numbers can change only at values
of $\om$ where either $b_0$ or $b_\pi$ in Eq.~\eqref{b0p} becomes equal to an 
integer. When that happens, Eq.~\eqref{uk1} becomes equal to $\pm I$ at either 
$k=0$ or $\pi$, and there is no gap to the Floquet eigenvalues at 
neighboring values of $k$. 


We have studied what happens for arbitrary (not necessarily positive)
values of $\ga$, $\De$, $c_0$, non-integer values of $2c_1 /\pi$, and $\om$.
The general result is as follows. Assuming that $b_{0/\pi}$
are not integers, we consider all the integers lying
between $b_0$ and $b_\pi$. Of these, let $n_e^>$ ($n_o^>$) and $n_e^<$
($n_o^<$) respectively denote the numbers of even (odd) integers which are
greater than and less than $2c_1 /\pi$. Then the numbers $N_+$ and $N_-$ of
modes at each end of a chain with Floquet eigenvalues $+1$ and $-1$ are given 
by $N_+ ~=~ |n_e^> - n_e^< |$ and $N_- ~=~ |n_o^> - n_o^< |$. (We will present
an explicit proof of this in Sec. V B for a special choice of parameters).
We also find that the winding number $W$ is given by $|W| = |n_e^> - n_e^< + 
n_o^> - n_o^<|$. Hence $|W|$ is generally {\it not} equal to the total number
of modes, $N_+ + N_-$, at each end of a chain (although
$|W| - (N_+ + N_-)$ is always an even integer). In Table I, we list the
values of $N_+$, $N_-$ and $|W|$ versus $\om$ for a 200-site system
with $\ga = 1$, $\De = -1$, $c_0 = 0.5$ and $c_1 = 0.2$.
In this case $c_0 + \ga > 0$, $c_0 - \ga < 0$ and $0 < 2c_1/\pi < 1$. Hence
$n_e^< , ~n_o^< \ne 0$ and $|W| \ne N_+ + N_-$ in general. In all cases, the
values of $N_+$, $N_-$ and $|W|$ obtained numerically and from 
Eq.~\eqref{heff2} match those obtained using $b_0$ and $b_\pi$ in 
Eq.~\eqref{b0p}. 

\begin{table}[htb]
\begin{center} \begin{tabular}{|c|c|c|c|c|c|c|c|c|c|c|c|c|c|c|c|c|c|c|}
\hline
$\om$ & 1 & 2 & 3 & 4 & 5 & 6 & 7 & 8 & 9 & 10 & 11 & 12 & 13 & 14 & 15 &
16 & 17 & 18 \\
\hline
$N_+$ & 2 & 0 & 0 & 1 & 1 & 1 & 1 & 1 & 1 & 1 & 1 & 1 & 1 & 1 & 1 &
0 & 0 & 0 \\
\hline
$N_-$ & 2 & 2 & 1 & 1 & 1 & 1 & 0 & 0 & 0 & 0 & 0 & 0 & 0 & 0 & 0 &
0 & 0 & 0 \\
\hline
$|W|$ & 4 & 2 & 1 & 0 & 0 & 0 & 1 & 1 & 1 & 1 & 1 & 1 & 1 & 1 & 1 &
0 & 0 & 0 \\
\hline
\end{tabular} \end{center}
\caption{Values of $N_+$, $N_-$ and $|W|$ versus $\om$ for $\ga = 1$, $\De 
= -1$, $c_0 = 0.5$ and $c_1 = 0.2$. $|W| \ne N_+ + N_-$ for $4 \le
\om \le 6$.} \label{tab2} \end{table}

In the limit $\om \to 0$, we can show from Eq.~\eqref{b0p} that the number 
of Majorana end modes diverges asymptotically as $8|\ga|/\om$ if $c_0 \pm 
\ga$ have the same sign and as $8|c_0|/\om$ if $c_0 \pm \ga$ have opposite 
signs. We see in Fig.~\ref{flo5}, particularly in the inset, that the number 
of end modes does diverge as $8/\om$ (recall that we have set $\ga =1$).

\subsection{Analytical results for Majorana end modes in a special case}

For the case $\De = -\ga$ and $c_0 = 0$, it turns out that we can
analytically find the wave functions of the Majorana end modes. Further, we
can explicitly prove that the number of Majorana modes is indeed governed by
the quantities $b_0$, $b_\pi$ and $2c_1/\pi$ as discussed above.

We consider a semi-infinite chain in which
$n$ goes from 1 to $\infty$ in Eq.~\eqref{ham1}; we will only discuss
the Majorana modes at the left end of this chain. As discussed above,
the Floquet operator which performs a time evolution for one time period
$T=2\pi/\om$ consists of a symmetrized product of three steps.
The first step evolves from time
$t=0$ to $t=\ep$ (where $\ep$ denotes an infinitesimal quantity),
the second step evolves from $t=\ep$ to $t=T- \ep$, and the third
step evolves from $t=T- \ep$ to $t=T$. At all times, the Heisenberg
operators $a_n (t)$ satisfy the equations $da_n(t)/dt = i [H(t),a_n (t)]$.
The first step corresponds to a Hamiltonian
\beq H_1 ~=~ \frac{ic_1}{2} ~\de (t - \frac{\ep}{2}) ~
\sum_{n=1}^\infty ~a_{2n-1} a_{2n}. \eeq
This gives
\bea a_{2n-1} (\ep) &=& a_{2n-1} (0) ~\cos c_1 ~+~ a_{2n} (0) ~\sin c_1,
\non \\
a_{2n} (\ep) &=& a_{2n} (0) ~\cos c_1 ~-~ a_{2n-1} (0) ~\sin c_1, \eea
for all $n \ge 1$.
The second step corresponds to the Hamiltonian
\beq H_0 ~=~ i \ga ~\sum_{n=1}^\infty ~a_{2n} a_{2n+1} \label{hmid} \eeq
for $\De = - \ga$ and $c_0 = 0$.
(The simple form in Eq.~\eqref{hmid} is a special feature of this
particular choice of $\ga$, $\De$ and $c_0$. For any other choice of these
parameters, the Hamiltonian would not decompose into terms involving pairs
of different Majorana operators, and the time evolution in this step would
not have a simple form). Eq.~\eqref{hmid} gives
\bea a_{2n} (T-\ep) &=& a_{2n} (\ep) ~\cos (2\ga T) ~+~ a_{2n+1}
(\ep) ~\sin (2\ga T), \non \\
a_{2n+1} (T-\ep) &=& a_{2n+1} (\ep) ~\cos (2\ga T) ~-~ a_{2n}
(\ep) ~\sin (2\ga T), \non \\
&& \eea
for all $n \ge 1$. Note that $a_1 (t)$ does not evolve in this step as $H_0$
does not contain $a_1$; hence $a_1 (T-\ep) = a_1 (\ep)$.
Finally, the third step corresponds to the Hamiltonian
\beq H_1 ~=~ \frac{ic_1}{2} ~\de (t - T + \frac{\ep}{2}) ~
\sum_{n=1}^\infty ~a_{2n-1} a_{2n}, \eeq
which gives
\bea a_{2n-1} (T) &=& a_{2n-1} (T - \ep) ~\cos c_1 ~+~ a_{2n} (T
- \ep)~\sin c_1,
\non \\
a_{2n} (T) &=& a_{2n} (T - \ep) ~\cos c_1 ~-~ a_{2n-1} (T- \ep) ~
\sin c_1. \non \\
&& \eea

We now discover that the equations above have two solutions for Majorana
end modes which correspond to Floquet eigenvalues being equal to $+1$ and 
$-1$, i.e., with $a_n (T) = \pm a_n (0)$ respectively for all $n \ge 1$. \\
\noi (i) For eigenvalues equal to $+1$, we find an unnormalized solution
of the form
\beq a_{2n-1} (0) = [\tan c_1 \cot (\ga T)]^n ~~~{\rm and}~~~ a_{2n} (0)
= 0 \label{evplus} \eeq
for all $n \ge 1$. \\
\noi (ii) For eigenvalues equal to $-1$, we find a solution of the form
\beq a_{2n-1} (0) = 0 ~~~{\rm and}~~~ a_{2n} (0) = [- \cot c_1 \cot (\ga
T)]^n \label{evminus} \eeq
for all $n \ge 1$. \\
We see that the wave function $a_n (0)$ is real, and the probability 
$|a_n (0)|^2$ has a very simple structure; depending on the Floquet 
eigenvalue, it vanishes for all odd $n$ or all even $n$, while for the 
other values of $n$ it decreases exponentially as $n$ increases.

Eqs.~(\ref{evplus}-\ref{evminus}) imply that Majorana
end modes appear or disappear when $|\tan c_1 \cot (\ga T)|$ or $|\cot c_1
\cot (\ga T)|$ becomes equal to 1. These are precisely the same conditions as
$b_0$ or $b_\pi$ in Eq.~\eqref{b0p} becoming equal to an integer $n$, with
Floquet eigenvalue equal to $(-1)^n$. We can also explicitly confirm the 
following result stated above. Namely, we consider all the integers lying
between $b_0$ and $b_\pi$, assuming that $b_0$, $b_\pi$ and $2c_1 /\pi$ are 
not integers. Of these, let $n_e^>$ ($n_o^>$) and $n_e^<$
($n_o^<$) respectively denote the numbers of even (odd) integers which are
greater than and less than $2c_1 /\pi$. Then the numbers $N_+$ and $N_-$ of
Majorana modes at the left of the chain with Floquet eigenvalue equal to
$+1$ and $-1$ are given by $N_+ ~=~ |n_e^> - n_e^< |$ and $N_- ~=~ |n_o^> - 
n_o^< |$. Interestingly, we find that $N_\pm$ can only be equal to 0 or 1
in this case.

\subsection{Effect of Time-Reversal Symmetry Breaking}
\begin{figure}[htb] \ig[width=3.4in]{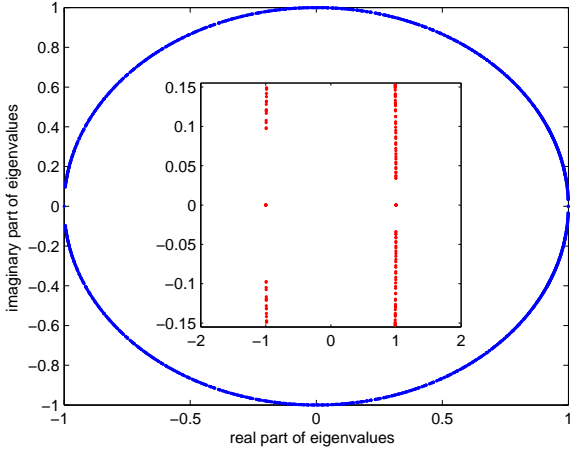}
\caption[]{(Color online) Floquet eigenvalues close to $\pm 1$ for a 1000-site 
system with 
$\ga =1$, $\De = -1$, $c_0=2.5$, and a $\de$-function kick with $c_1=0.2$ at 
$t=0$ which is repeated with a time period $T=2\pi$. For this time-reversal 
symmetric case, there are four eigenvalues at exactly $+1$ and $-1$ each, 
separated by a gap from all other eigenvalues; these are shown more clearly 
in the inset.} \label{flo7} \end{figure}

\begin{figure}[htb] \ig[width=3.4in]{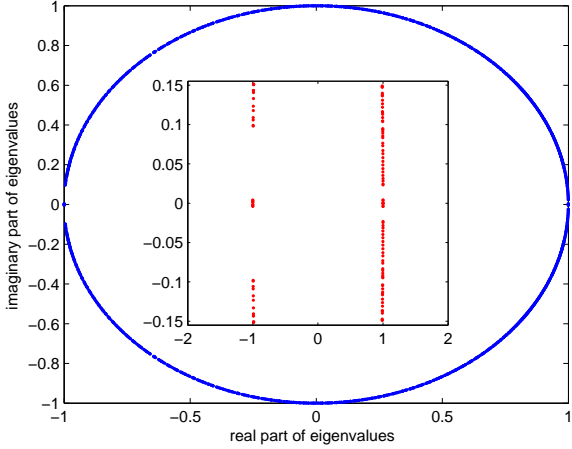}
\caption[]{(Color online) Floquet eigenvalues close to $\pm 1$ for a 
1000-site system with 
$\ga =1$, $\De = -1$, $c_0=2.5$, and two $\de$-function kicks with $c_1=0.2$ 
and $c_2= 0.1$ at $t=0$ and $T/4$ which are repeated with a time period $T=2
\pi$. For this case with no time-reversal symmetry, there are four eigenvalues
close to but not exactly at $+1$ and $-1$, separated by a gap from all other 
eigenvalues; these are shown more clearly in the inset.} 
\label{flo8} \end{figure}

Given a periodically driven time-reversal symmetric system which has Majorana 
end modes (namely, modes with real eigenvectors and Floquet eigenvalues equal 
to $\pm 1$ which are separated from all other eigenvalues by a gap), we may 
ask what would happen if we add small terms which break time-reversal 
symmetry. We discover that the end modes persist and their Floquet 
eigenvalues continue to be separated from all other eigenvalues by a gap. 
However, the Floquet eigenvalues move slightly away from $\pm 1$ in complex 
conjugate pairs, and the eigenvectors become complex; hence they can no
longer be called Majorana modes. This is illustrated in Fig.~\ref{flo7} 
which shows the Floquet eigenvalues for the modes at each end of the chain
for the time-reversal symmetric case given in Eq.~\eqref{ht1}, while 
Fig.~\ref{flo8} shows the Floquet eigenvalues for a case with 
\bea \mu_n (t) &=& c_0 ~+~ c_1 \sum_{n=-\infty}^\infty \de (t - nT) \non \\
& & +~ c_2 \sum_{n=-\infty}^\infty \de (t - \frac{T}{4} - nT), \eea
which breaks time-reversal symmetry. Although we cannot clearly see from 
Fig.~\ref{flo8} that the Floquet eigenvalues of the end modes have moved 
away from $\pm 1$, we have checked numerically that this is so. For a 
2800-site system (this is a large enough system size that there is no mixing
between the two ends), we find that at each end, the Floquet eigenvalues near 
$-1$ are given by $-1 \pm 0.0015 i$ and $-1 \pm 0.0037 i$, and the eigenvalues
near $+1$ are given by $1 \pm 0.0007 i$ and $1 \pm 0.0040 i$.

\section{Periodic $\de$-function Kicks in Hopping and Superconducting Terms}
In this section, we will briefly discuss the case where the hopping and
superconducting terms in Eq.~\eqref{ham1} are given $\de$-function kicks 
periodically in time. We will again show that this too can produce Majorana
end modes. In particular, we find that there is a Majorana mode at
each end of a chain even in the limit of very large driving frequency $\om$;
this is in contrast to the case of periodic $\de$-function kicks in the 
chemical potential where there is an upper limit on $\om$ beyond which there 
are no Majorana modes. We will limit our discussion to some observations on 
the Floquet operator and the winding number; we will not consider the 
possibility of a second topological invariant here.

We consider the case where the chemical potential $\mu$ is independent
of time, while 
\beq \ga ~=~ - \De ~=~ \ga_0 ~+~ \ga_1 ~\sum_{n=-\infty}^\infty \de (t - nT).
\eeq
(As mentioned earlier, this corresponds to an Ising model in a transverse
magnetic field, where the Ising interaction $J_x$ is given periodic 
$\de$-function kicks while the magnetic field does not vary with time).
Eqs.~\eqref{hk} and \eqref{flo2x} then imply that 
\bea U_k (T,0) &=& e^{-i \ga_1 (\cos k ~\tau^z - \sin k ~\tau^y)} \non \\
& & \times ~e^{-i2T[(\ga_0 \cos k - \mu) \tau^z - \ga_0 \sin k ~\tau^y]} 
\non \\
& & \times ~e^{-i \ga_1 (\cos k ~\tau^z - \sin k ~\tau^y)}. \label{uk2} \eea

Eq.~\eqref{uk2} implies that in the limit $\om \to \infty$, i.e., $T \to 0$,
$U_k (T,0) = e^{-i(a_{2,k} \tau^y + a_{2,k} \tau^z)}$, where 
\beq a_{2,k} ~=~ - 2 \ga_1 \sin k ~~~~{\rm and}~~~~ a_{3,k} ~=~ 2 \ga_1 \cos 
k. \eeq
As $k$ goes from $-\pi$ to $\pi$, this generates a closed curve with winding 
number $+1$. This implies that there will be one Majorana mode at each end of 
the chain when $\om \to \infty$. Numerically, we find that this is indeed the 
case. We will now prove this analytically for a special set of parameters 
following a procedure similar to the one followed in Sec. V B.

We consider the case $\ga_0 = 0$. Considering only the left end of
the chain starting from $n=1$ and assuming some initial values of the
Heisenberg operators $a_n (0)$, we can successively find 
$a_n (\ep)$, $a_n (T - \ep)$ and $a_n (T)$ using three sets of 
evolution equations,
\bea a_{2n+1} (\ep) &=& a_{2n+1} (0) ~\cos \ga_1 ~-~ a_{2n} (0) ~\sin \ga_1, 
\non \\
a_{2n} (\ep) &=& a_{2n} (0) ~\cos \ga_1 ~+~ a_{2n+1} (0) ~\sin \ga_1, 
\label{time1} \eea
\bea a_{2n-1} (T-\ep) &=& a_{2n-1} (\ep) ~\cos (2\mu T) ~+~ a_{2n} (\ep)~
\sin (2\mu T), \non \\
a_{2n} (T-\ep) &=& a_{2n} (\ep) ~\cos (2\mu T) ~-~ a_{2n-1} (\ep) ~\sin (2
\mu T), \non \\
&& \label{time2} \eea
and
\bea a_{2n+1} (T) &=& a_{2n+1} (T-\ep) ~\cos \ga_1 ~-~ a_{2n} (T-\ep) ~\sin 
\ga_1, \non \\
a_{2n} (T) &=& a_{2n} (T-\ep) ~\cos \ga_1 ~+~ a_{2n+1} (T-\ep) ~\sin \ga_1. 
\non \\
&& \label{time3} \eea
Eqs.~(\ref{time1}-\ref{time3}) are valid for all $n \ge 1$. Note that $a_1$ 
does not evolve at all from $t=0$ to $t=\ep$ and again from $t=T-\ep$ to $t=T$.

We then discover that the above equations have two kinds of solutions. \\ 
\noi (i) For Floquet eigenvalue equal to $+1$, i.e., $a_n (T) = a_n (0)$ for 
all $n$, we find an unnormalized solution of the form $a_1 (0) = 1$, while
$a_{2n+1} = [\tan( \mu T) \cot \ga_1]^n /\cos \ga_1$ and $a_{2n} (0) 
= 0$ for all $n \ge 1$. This solution exists if $|\tan( \mu T) \cot \ga_1| < 
1$. In the limit $T \to 0$, it reduces to $a_1 (0) = 1$ and all other $a_n 
(0) = 0$. \\
\noi (i) For Floquet eigenvalue equal to $-1$, i.e., $a_n (T) = - a_n (0)$, 
we find an unnormalized solution of the form $a_1 (0) = 1$, while $a_{2n+1} = 
[- \cot( \mu T) \cot \ga_1]^n /\cos (\ga_1)$ and $a_{2n} (0) = 0$ for all 
$n \ge 1$. This solution exists if $|\cot( \mu T) \cot \ga_1| < 1$.

\section{Simple Harmonic Variation of Chemical Potential with Time}
In this section, we discuss the case where the chemical potential varies 
harmonically with $t$. Namely, the chemical potential in Eq.~\eqref{ham2} 
takes the form 
\beq \mu (t) ~=~ c_0 ~+~ c_1 \cos (\om t + \phi). \label{ht2} \eeq
The Floquet operator can be written as the time-ordered product
\beq U(T,0) = {\cal T} e^{4\int_0^T dt M(t)}, \eeq
where $M(t)$ is an antisymmetric matrix with the non-zero elements
\bea (M)_{2n+1,2n} &=& - ~(M)_{2n,2n+1} ~=~ - \frac{1}{4} (\ga - \De), \non \\
(M)_{2n-1,2n+2} &=& - ~(M)_{2n+2,2n-1} ~=~ - \frac{1}{4} (\ga + \De), \non \\
(M)_{2n-1,2n} &=& - ~(M)_{2n,2n-1} \non \\
&=& \frac{1}{2} [c_0 + c_1 \cos (\om t + \phi)]. \label{mt2} \eea 
Unlike the case of the periodic $\de$-function kick, the Floquet operator is
no longer a product of only two or three operators; it has to be computed by 
dividing the time-period $T$ into a large number of time steps of size 
$\De t$ each, and then multiplying $T/\De t$ operators in a time-ordered 
way. Finally, we have to check that the 
results do not change significantly once $\De t$ has been made sufficiently
small. Hence this problem takes much more computational time. For the same
reason, a numerical calculation of the Floquet operator $U_k (T,0)$ takes 
more time here than the corresponding expression given in Eq.~\eqref{uk1} for 
a periodic $\de$-function kick. We will not consider the existence of 
topological invariants here.

Having computed the operator $U(T,0)$, where $T=2\pi/\om$, we again find all 
the eigenvalues and eigenvectors of $U(T,0)$ which are also eigenvectors of 
the parity operator $\cal P$. As functions of the parameters $\ga$, $\De$, 
$c_0$, $c_1$, $\phi$, $\om$ and $N$, we find that the qualitative features of 
the Majorana end modes that we find are similar to the case of the 
periodic $\de$-function kicks. As before, we find that end modes 
can appear even when the chemical potential places the corresponding 
time-independent system in a non-topological phase at all times $t$.

The effect of the phase $\phi$ in Eqs.~\eqref{ht2} and \eqref{mt2} turns 
out to be interesting. The Floquet operator $U(T,0)$, now denoted 
by $U_\phi (T,0)$, clearly depends on $\phi$. However, we can show that the 
eigenvalues of $U_\phi (T,0)$ are independent of $\phi$~\cite{soori}. To see 
this, note that a shift in the phase $\phi$ by an amount $\de$ is equivalent
to a shift in time by the amount $\de/\om$. Hence 
\bea U_\phi (T,0) &=& U_0 (T+\phi/\om, \phi/\om) \non \\
&=& U_0 (T+\phi/\om, T) U_0 (T, \phi/\om) \non \\
&=& U_0 (T+\phi/\om, T) U_0 (T, 0) U_0^{-1} (\phi/\om, 0) \non \\
&=& U_0 (\phi/\om, 0) U_0 (T, 0) U_0^{-1} (\phi/\om, 0), \label{uphi} \eea 
where we have used the fact that $U_0 (T+\phi/\om, T)= U_0 (\phi/\om, 0)$.
Eq.~\eqref{uphi} shows that $U_\phi (T,0)$ is related to $U_0 (T, 0)$ by a
unitary transformation involving $U_0 (\phi/\om, 0)$; hence they have the 
same eigenvalues while their eigenvectors are related by the same unitary 
transformation. This also implies that studying how an eigenvector 
corresponding to a particular eigenvalue of $U_\phi (T,0)$ changes with 
$\phi$ is equivalent to studying how that eigenvector changes with time under
evolution with $U_0 (t=\phi/\om ,0)$. 

\begin{figure}[htb] \ig[width=3.4in]{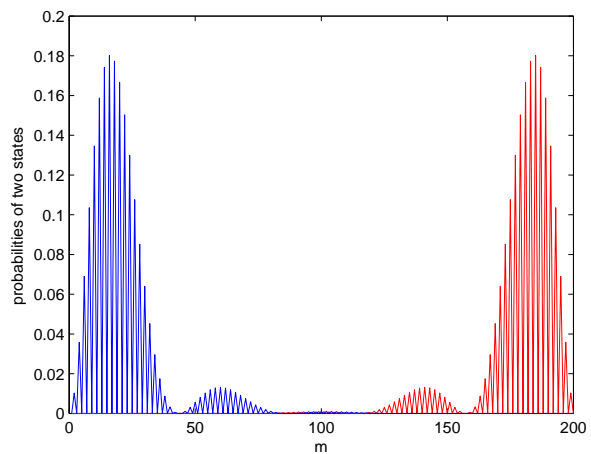}
\caption[]{(Color online) Plot of the probabilities of the two end states for 
a 100-site 
system with $\ga = 1$, $\De = -1$, $c_0 = 2.5$, and a simple harmonic driving
with $c_1 = 0.3$, $\om = 14$ and $\phi = 0$.} \label{flo9} \end{figure}

\begin{figure}[htb] \ig[width=3.4in]{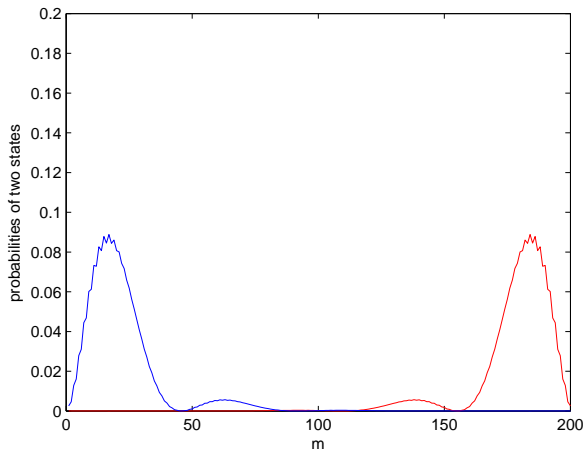}
\caption[]{(Color online) Plot of the probabilities of the two end states 
for a 100-site 
system with $\ga = 1$, $\De = -1$, $c_0 = 2.5$, and a simple harmonic driving
with $c_1 = 0.3$, $\om = 14$ and $\phi = \pi/2$.} \label{flo10} \end{figure}

The effect of a phase change on the wave functions of the Majorana
end modes can sometimes be quite dramatic. We consider a 100-site
system with $\ga = 1$, $\De = -1$, $c_0 = 2.5$, $c_1 = 0.3$ and $\om = 14$.
Figs.~\ref{flo9} and \ref{flo10} show the probabilities of the two end modes 
for $\phi = 0$ and $\pi/2$ respectively; equivalently, we can think of these
figures as showing the effect of evolving the end modes by a time of $T/4$.
We see that the detailed form of the end mode wave functions are quite
different in the two cases. For $\phi =0$, the wave function $a_m$ of the
mode at the left (right) end is non-zero only if $m$ is even (odd).
For $\phi = \pi/2$, both end modes have wave functions in which $a_m$ is
non-zero for both even and odd values of $m$.

\section{Effects of electron-phonon interactions and noise on Majorana end
modes}
An important question relevant to the experimental detection of Majorana end
modes generated by periodic driving is whether such modes are stable under
perturbations which do not have the same periodicity as the driving term.
For instance, at finite temperature, there will be phonons with a range of
frequencies $\om'$, and we may be interested in effect of electron-phonon
interactions on the Majorana end modes. We may also be interested in the
effect of a random noise in some of the parameters in the Hamiltonian.
We will discuss both these questions here.

Given a Majorana mode produced by driving with a frequency $\om$, let us
define the quasienergy gap as $\De E = \om \De \ta /(2\pi)$, where $\De \ta$
is the gap between the Floquet eigenvalues of the bulk modes 
(which are of the form $e^{i\ta}$)
and the Majorana mode (which must have $e^{i\ta} = \pm 1$, i.e., $\ta = 0$ or
$\pi$). It has been shown in Ref. \onlinecite{kita2} that the Majorana mode
will survive if the phonon frequencies $\om'$ (which are mainly governed by
the temperature) are much smaller than the gap $\De E$
and the driving frequency $\om$ is much larger than both $\om'$ and
the bandwidth. The basic argument for this result is that the driving
with frequency $\om$ and an interaction between an electron and a phonon
with frequency $\om'$ can combine to produce transitions between two states
whose energies differ by $\om' + n \om$, where $n$ is an integer. (We are
assuming here that the electron-phonon interaction is small so that only
one-phonon processes are important). If $n=0$,
then we cannot have a transition between the Majorana mode and a bulk mode
if $\om' \ll \De E$. On the other hand, if $n \ne 0$, then
$\om' + n \om$ is much larger than the bandwidth; then there is no bulk mode
available to which we can make a transition from the Majorana mode.

Applied to our model, the argument outline above implies that if $\om$ is much
larger than the bandwidth of the time-independent part of the Hamiltonian
(this is equal to $4 |\ga + c_0|$ as we can show using Eq.~\eqref{disp}), a 
Majorana mode will survive if the phonon frequencies are much smaller than 
the corresponding quasienergy gap $\De E$. Conversely, if $\om$ is of the 
order of or smaller than the bandwidth, then the Majorana mode may not be 
stable against electron-phonon interactions. The large number of Majorana 
modes that we found in Sec. V A for very small values of $\om$ may therefore 
not be stable against electron-phonon interactions.

We have numerically also studied what happens when the chemical potential has
a term which is uniform in space but varies randomly in time; in addition,
the chemical potential is given periodic $\de$-function kicks. We compute 
the Floquet operator by dividing the total time $\bar T$ into a large number 
of steps (of size $\De \bar T$ each) and multiplying the time evolution
operators over all the steps in a time-ordered way. We consider a chain with
$\ga =1$, $\De = -1$, $c_0 = 2.5$, $c_1 = 0.2$, with a range of system sizes
from 200 to 1000 and a range of frequencies $\om$ from 1 to 16. To study
the effect of
noise, we add a term to the chemical potential $\mu (t)$ which is of the form
$r p(t)$, where $p(t)$ is a random variable which is uniformly distributed
from $-1$ to $1$ and is uncorrelated at different times (this is achieved by
choosing $p(t)$ to be a different random number at each time step of our
numerical calculations), and $r$ is the coefficient of the random term. We
have studied the effect of the noise over a total time $\bar T$ ranging from
$T$ to $11 T$, where $T = 2\pi/\om$. (This implies that our noise has a period
varying from $T$ to $11 T$, rather than being truly aperiodic). We find
that for $4 \le \om \le 16$, the Majorana end modes survive up to a value of
$r$ which is about $0.3$. For smaller values of $\om = 2, 3$, the
Majorana modes survive up to a value of $r$ of about $0.05$,
while for $\om = 1$, they survive up to $r$ of about $0.025$. The critical
value of $r$ varies somewhat from one run to another as is expected for a
random noise. (We have not studied how $r$ depends on the quasienergy gap
$\om \De \ta /(2\pi)$; note that this gap also generally decreases as $\om$
decreases). To conclude, a noise in the chemical potential does not destroy
the Majorana modes if the strength of the noise is less than some value
which decreases with the driving frequency $\om$.

We note that electron-phonon interactions and noise do not have the same
effects in our system. The random noise that we have considered contains
terms with a very large number of frequencies ranging from $2\pi/\bar T$ to
$2\pi /\De \bar T$, and all these terms interact with the electrons.
On the other hand, we have only considered processes in which only one
phonon interacts with the electrons and each phonon has a single
frequency $\om'$. The electron-phonon interactions and noise therefore
affect the Majorana modes in different ways.

\section{Conclusions}

In this work we have shown that periodic driving of a one-dimensional
model of electrons with $p$-wave superconductivity or a spin-1/2 $XY$ chain
in a transverse magnetic field
can generate Majorana modes at the ends if we have a large and
open system which is time-reversal symmetric. To simplify the calculations, we 
have mainly studied the case in which the chemical potential of the electrons
(or the transverse magnetic field in the spin language) is given a periodic
$\de$-function kick. However, similar results are found when the chemical
potential (or magnetic field) is driven in a simple harmonic way, or when
the hopping and superconducting terms are given periodic $\de$-function kicks. 

The Majorana end modes exist only for very large system sizes and have three 
characteristic features: the Floquet eigenvalues are exactly equal to $\pm 1$,
they are separated from all the other eigenvalues by a finite gap, and the 
wave functions are real. If the system is not time-reversal symmetric, we find
that there may still be end modes whose eigenvalues are separated from all the
others by a finite gap may; however, the eigenvalues are no longer exactly at 
$\pm 1$, and the wave functions are not real. Hence these cannot be called
Majorana modes.

In analogy with the known topological invariants which predict the number 
of zero energy Majorana end modes for a system with a time-independent 
Hamiltonian, we have studied if the driven system has topological invariants 
which can correctly predict the number of end modes. We have shown that 
there are two topological invariants which work for a wide range of the
driving frequency $\om$ for the case of the periodic $\de$-function kick.
The first invariant is a winding number which is similar in form to the 
topological invariant for a time-independent Hamiltonian with time-reversal 
symmetry; this invariant sometimes, but not always, gives the total number 
of end modes. The second invariant is superior in that it separately gives 
us the numbers of end modes with Floquet eigenvalues equal to $+1$ and $-1$ 
for all values of the parameters. The second invariant also gives us a simple 
condition which can predict the values of $\om$ at which end modes appear or 
disappear. 

We have studied the effects of some experimentally relevant perturbations 
such as electron-phonon interactions and a random noise on the Majorana 
end modes. We generally find that the Majorana modes become more robust as 
the driving frequency $\om$ increases.

Recently there has been considerable excitement over claims of the detection 
of Majorana modes in semiconducting/superconducting 
nanowires~\cite{kouwenhoven,deng,rokhinson,das,finck} following some 
theoretical proposals~\cite{lutchyn1,oreg,alicea,stanescu}. A zero bias peak 
has been observed in the tunneling conductance into one end of the nanowire, 
and it has been suggested that this is the signature of a Majorana end mode. 
Our results can be tested in similar systems by applying a gate voltage to the 
nanowire which varies periodically in time in some way. One would like to 
see if such a time-dependent gate voltage can give rise to a zero bias peak;
this has recently been studied in Ref.~\onlinecite{kundu}. An important 
question which needs to be investigated in this context is how the Majorana 
end modes appear in the 
steady state after the oscillatory part of the gate voltage is switched 
on. This would require a treatment of various relaxation mechanisms which
may be present in the system~\cite{lind1}. Finally, the effects that disorder 
in the chemical potential~\cite{motrunich,brouwer1,lobos,cook,pedrocchi} and 
electron-electron interactions~\cite{ganga,lobos,lutchyn2,fidkowski2} have on 
the Majorana end modes also need to be examined.

\section*{Acknowledgments}
For financial support, M.T. and A.D. thank CSIR, India and D.S. thanks DST, 
India for Project No. SR/S2/JCB-44/2010.

\end{document}